\documentclass[conference,letterpaper]{IEEEtran}

\usepackage[utf8]{inputenc} 
\usepackage[T1]{fontenc}
\usepackage{url}
\usepackage{ifthen}
\usepackage{cite}
\usepackage[cmex10]{amsmath} 
\usepackage{amsfonts, amsthm, amssymb}
\usepackage{algorithm} 
\usepackage{algpseudocode} 
\usepackage{array}
\usepackage[caption=false,font=normalsize,labelfont=sf,textfont=sf]{subfig}
\usepackage{textcomp}
\usepackage{stfloats}
\usepackage{url}
\usepackage{verbatim}
\usepackage{graphicx}
\usepackage{cite}
\usepackage{xcolor}

\normalsize

\newtheoremstyle{bfnote}%
{}{}
{\itshape}{}
{\bfseries}{.}
{ }{\thmname{#1}\thmnumber{ #2}\thmnote{ (#3)}}
\theoremstyle{bfnote}

\newtheorem{theorem}{Theorem}

\newtheorem{definition}{Definition}
\newtheorem{proposition}{Proposition}
\newtheorem{axiom}{Axiom}
\DeclareMathOperator*{\argmax}{arg\,max}

\begin{document}
\title{On The Theory of Semantic Information and Communication for Logical Inference} 

\author{%
  \IEEEauthorblockN{Ahmet Faruk Saz, Siheng Xiong and Faramarz Fekri}
  \IEEEauthorblockA{Georgia Institute of Technology\\
                    Atlanta, GA, USA\\
                    Email: \{asaz3, sxiong45\}@gatech.edu, faramarz.fekri@ece.gatech.edu}
}

\maketitle

\begin{abstract}

First-Order Logic (FOL), also called first-order predicate calculus, is a formal language that provides a framework to comprehensively represent a world and its present state, including all of its entities, attributes, and complex interrelations, irrespective of their physical modality (e.g., text, image, or sensor data). Grounded in this universal representation, this paper develops a mathematical theory for semantic information and communication tailored to tasks involving logical reasoning and inference. For semantic communication, our framework distinguishes between two fundamental components: the physical cost of transmitting symbols of the FOL language and the logical content those symbols represent. A calibrated measure for semantic content is proposed, which allows for the consistent comparison of information value across different logical systems. This measure quantifies the degree to which a message reduces uncertainty about the true state of the world. Building on this measure, semantic entropy, conditional and mutual information metrics are defined. These metrics are then used to formulate optimizable objectives for semantic communication, designed to preserve the information most relevant for logical reasoning task at the receiver while adhering to a transmission budget. The framework's operational value is demonstrated through experiments in semantic compression, where the proposed objectives are used to manage the trade-off between transmission cost and the preservation of logical content; and deductive inference, where increasing world-state awareness improves deduction performance.
\end{abstract}


\vspace{-5pt}
\section{Introduction}

Semantic communication is emerging as a critical paradigm to overcome the inefficiencies of conventional systems struggling with AI-driven, data-intensive applications ~\cite{gunduz2022semantic}. This is especially true for symbolic tasks and logical inference—processes that involve reasoning over structured knowledge to enable verifiable, explainable, and complex decision-making in cyber-physical systems, advanced IoT ecosystems and edge-intelligent networks. In these systems, the quality of high-stakes decisions and the accuracy of inference depend directly on receiving a timely and comprehensive representation of the current state of the world. However, transmitting raw sensory data to capture this state is often infeasible due to bandwidth and latency limitations. Then, the goal is to transmit a compact yet informative semantic representation of the world state. This representation may take various forms, such as a distilled logical conclusion (e.g., $Brake(Immediately)$ derived from fusing multiple sensor inputs), a prioritized subset of FOL statements, or an efficiently encoded representation. This approach ensures that even when the communication budget is limited, the knowledge most vital for effective, real-time decision-making in applications like industrial fault detection or vehicular safety is successfully prioritized and transmitted~\cite{qin2022semantic}.

Despite this potential, a central challenge remains: the development of a high-level mathematically grounded theory to guide the design of semantic communication systems for symbolic inference~\cite{kountouris2021role}. This work establishes such a formal, interpretable foundation by addressing a core problem identified by Weaver: quantifying "how precisely the transmitted symbols convey the desired meaning"~\cite{shannon1949mathematical}. To achieve this, we distinguish between the subjective, goal-oriented aspects of communication (pragmatics) and the quasi-objective, logical content of a message (semantics). Our framework focuses exclusively on the latter, proposing a quantitative theory for logical inference inspired by the pioneering work of Carnap and Bar-Hillel on logical probability and information content~\cite{carnap1952outline}. We adopt First-Order Logic (FOL) as the formal language to represent this semantic content, as it offers a powerful and universally interpretable structure~\cite{TILR,TILP,TEILP}. To quantify information within this system, our approach draws on the semiotic distinction between the signifier (the physical symbols or bits representing a predicate or entity) and the signified (the conceptual meaning). The semantic value of a sign—the union of a signifier and its signified—is thus determined by its relationship to other signs within the logical structure, allowing our framework to analyze the information content of semantic transmissions.

To operationalize this theory, our methodology is grounded in the inductive logic of Carnap and Hintikka. We begin by distinguishing between two fundamental types of semantic uncertainty—over the message itself and its underlying content—and formally characterize both using an inductive probability framework built on FOL. This approach yields calibrated measures of semantic information, allowing for the direct comparison of content-informativeness even between systems using disparate FOL-based ontologies. To manage the computational intractability of operating over vast logical spaces, we introduce a practical method of dynamically constructing evidential sub-languages. Building on this, we utilize Hintikkan conditional and mutual content-information measures to formulate distinct, optimizable rate-distortion objectives for lossy semantic compression. The performance of this framework is then validated through experiments that measure compression efficiency and the fidelity of the retained semantic content over a custom dataset and deductive inference performance of autonomous vehicles in FOL-based urban city simulator LogiCity.

\section{Related Works}

Research in semantic communication, motivated by Weaver's Level B, has advanced along two primary fronts. The first, rooted in deep learning, utilized autoencoder-based systems like DeepSC~\cite{xie2021deep}, DeepJSCC~\cite{bourtsoulatze2019deep} and functional compression~\cite{saidutta2021joint} to performs neural network-based efficient compression and transmission of meaningful content. This paradigm was later extended to task-oriented training for distributed transceivers~\cite{saidutta2021analog}, attention-based semantic filtering~\cite{yang2023task}, and multi-agent cooperative systems~\cite{xie2022task}. Concurrently, a second stream focuses on neuro-symbolic methods for more interpretable semantics, using knowledge graphs and large language models to extract and compress structured triplets~\cite{guo2023kg}. By integrating multi-modal tools like SAM and GPT~\cite{jiang2024semantic} and leveraging graph neural networks~\cite{nour2024semantic}, these approaches use generative AI to reconstruct rich content, achieving large compression ratios and improving semantic integrity.

\section{First-Order Language for Semantic Structures}

Our semantic communication framework is built upon a dyadic first-order language, $\mathcal{L}$, over a universe of discourse $\mathcal{U}$ containing a set of entities $\mathcal{E}$. The language's vocabulary consists of a finite set of $m$ dyadic predicates, $\{\mathcal{P}_1, \ldots, \mathcal{P}_m\}$. Atomic formulas of the form $\mathcal{P}_j(x, y)$, where $x, y \in \mathcal{E}$, represent relationships between entities. These are combined using logical connectives ($\land, \lor, \neg$) to form complex well-formed formulas that describe the state of the world.

To structure this language for probabilistic inference, we define a hierarchy of logical components based on the work of Carnap and Hintikka \cite{c1, c2, c3}.

\textbf{R-conjunctions:} An R-conjunction, denoted $R_i(x, y)$, provides a complete description of the directed relationship from entity $x$ to $y$. It is a conjunction of every predicate or its negation:
$$R_i(x, y) \triangleq (\pm)\mathcal{P}_1(x,y) \land \ldots \land (\pm)\mathcal{P}_m(x,y)$$
where $(\pm)$ indicates the presence or absence of negation. There are $2^m$ distinct R-conjunctions.

\textbf{Q-predicates:} A Q-predicate, $Q_k(x,y)$, captures the complete, potentially asymmetric, two-way relationship between a pair of entities. It is formed by the conjunction of two R-conjunctions, one for each direction:
$$Q_k(x, y) \triangleq R_i(x,y) \land R_j(y,x)$$
A Q-predicate thus describes the full relational state between $x$ and $y$. For example, the statement 'Alice likes Bob, but Bob does not like Alice' would correspond to a specific Q-predicate.

\textbf{Attributive Constituents:} An attributive constituent, $Ct_i(x)$, defines the "kind" of an entity $x$ by exhaustively specifying its pattern of relationships with all other entities. It asserts the complete set of Q-predicates that $x$ participates in:
\begin{align}
    Ct_{i}(x) \triangleq & (\exists y)Q_{i_1}(x,y) \land \ldots \land (\exists y)Q_{i_h}(x,y) \land \nonumber \\
    & (\forall y)\{Q_{i_1}(x,y) \lor \ldots \lor Q_{i_h}(x,y)\}
\end{align}
This formula states that entity $x$ engages in relationships of types $Q_{i_1}, \ldots, Q_{i_h}$ and no others.

\textbf{Constituents:} A constituent, $C_w$, provides a complete, holistic description of the entire universe $\mathcal{U}$ by stating exactly which kinds of entities exist. A constituent of width $w$ asserts the existence of precisely $w$ distinct kinds of entities and that no other kinds exist:
\begin{align}
    C_w \triangleq & (\exists x)Ct_{i_1}(x) \land \ldots \land (\exists x)Ct_{i_w}(x) \land \nonumber \\
    & (\forall x)(Ct_{i_1}(x) \lor \ldots \lor Ct_{i_w}(x))
\end{align}
Constituents are mutually exclusive and jointly exhaustive, representing the fundamental, indivisible semantic states of the language $\mathcal{L}$. A key property of this structure is that any formula (or "clause") $h$ expressible in $\mathcal{L}$ is logically equivalent to a finite disjunction of constituents \cite{c4, c5, c6, c7}:
$$h \equiv \bigvee_{i \in \mathcal{I}} C_i$$
This property establishes the set of constituents as a canonical basis for defining a probability distribution over all possible states of the world described by $\mathcal{L}$.

\section{Inductive Probability over Semantic Structures}

With constituents established as the semantic atoms of $\mathcal{L}$, we now assign them probabilities using the principles of inductive logic \cite{c8, c9, c10, c11}. The inductive probability of any formula $h$ is derived from a probability distribution over the constituents; given that $h \equiv \bigvee_{i \in \mathcal{I}} C_i$, its probability is the sum of the probabilities of the constituents in its disjunctive normal form.

The central concept in this framework is the degree of confirmation, $c(h, e)$, which quantifies the rational support that evidence $e$ provides for a hypothesis $h$. It is defined as a conditional probability:
$$q_e(h) \triangleq c(h, e) \triangleq p(h|e) = \frac{p(h \land e)}{p(e)}$$
Here, $p(\cdot)$ is the prior inductive probability distribution over the constituents, reflecting a state of knowledge based solely on the logical structure of the language before any empirical evidence is considered.

This single formulation serves as a robust engine for reasoning under uncertainty. The degree of confirmation $c(C_i, e)$ directly measures the rational belief that a specific constituent $C_i$ represents the true state of the world, where evidence $e$ can include both direct observations and received semantic information. More broadly, this enables various forms of inductive inference by quantifying the evidential support for diverse hypotheses ($h$). These include singular predictions (e.g., the likelihood of a future observation), universal generalizations (e.g., a law holding true for all entities), structural hypotheses about the universe's complete description, abductive explanations for observations, and even counterfactual claims. This framework moves beyond the strict entailment of deductive logic to a more flexible, evidence-based system of confirmation, recovering deduction as the special case where a degree of confirmation of one ($c(h, e) = 1$) signifies that the evidence logically entails the hypothesis ($e \vDash h$).

\section{Derivation of the Inductive Characteristic Function}

To compute inductive probabilities, we require a principled method for updating beliefs. Our framework's inductive characteristic function, which determines the predictive probability $p(\mathcal{A}_i | e)$ of an outcome $\mathcal{A}_i$ given evidence $e$, is derived from Bayesian principles that generalize Laplace's Rule of Succession.

Consider a process with $K$ possible, mutually exclusive outcomes $\mathcal{A} = \{ \mathcal{A}_1, \ldots, \mathcal{A}_K \}$. We model the unknown long-run frequencies of these outcomes, $(p_1, \ldots, p_K)$, as random variables governed by a probability distribution. We begin by assuming a symmetric Dirichlet prior, which is the conjugate prior for the multinomial likelihood:
\vspace{-5pt}
$$(p_1, \ldots, p_K) \sim \operatorname{Dirichlet}(\alpha, \ldots, \alpha), \quad \text{where } \alpha = \frac{\lambda}{K}$$
The parameter $\lambda > 0$ controls the strength of the prior belief; a smaller $\lambda$ reflects a weaker initial assumption that the outcomes are equiprobable.

Let the evidence $e$ consist of $n$ observations with counts $n_i$ for each outcome $\mathcal{A}_i$, such that $\sum_{i=1}^K n_i = n$. By Bayes' theorem, the posterior distribution over the frequencies is also a Dirichlet distribution:
$$(p_1, \ldots, p_K) | e \sim \operatorname{Dirichlet}(n_1 + \alpha, \ldots, n_K + \alpha)$$The predictive probability for the next outcome to be $\mathcal{A}_i$ is the expected value of the posterior marginal for $p_i$. This yields our characteristic function:$$p(\mathcal{A}_i | e) = \mathbb{E}[p_i | e] = \frac{n_i + \alpha}{n + K\alpha} = \frac{n_i + \lambda / K}{n + \lambda}$$
This function elegantly balances prior belief (controlled by $\lambda$) with empirical data (the counts $n_i$ and total $n$). In the limit of infinite evidence ($n \to \infty$), the law of large numbers ensures that the predictive probability converges to the true frequency, $p(\mathcal{A}_i | e) \to f_i$, guaranteeing consistency with classical statistics. Consequently, the expected information content of the predictions converges to the Shannon entropy of the true underlying distribution, $-\sum_i f_i \log f_i$.

\section{Computing the Inductive Posterior over Constituents}

The core of our probabilistic framework is the posterior distribution over constituents, $p(C_w | e)$, which quantifies the rational belief in each possible universal structure given the observed evidence $e$. Evidence is formally a clause specifying the Q-relations among a set of $n$ observed individuals. The posterior is computed via Bayes' theorem:
$$p(C_w | e) = \frac{p(C_w)p(e | C_w)}{\sum_{i} p(C_i)p(e | C_i)}$$
Following Hintikka's system, the prior $p(C_w)$ (reflecting initial belief in the universe's complexity) and the likelihood $p(e | C_w)$ are derived from the principles of inductive logic, with their full formulations involving complex Gamma functions [c10].

However, a direct application of this formalism is intractable. The state space of constituents is combinatorially vast, and real-world evidence $e'$ is almost always incomplete, violating the theoretical requirement for a complete specification of all relations. To overcome these hurdles, we employ an evidential sub-language approach \cite{Niiniluoto1973TheoreticalCA}. For any instance of incomplete evidence $e'$, we dynamically construct a sub-language $\mathcal{L}' \subset \mathcal{L}$ that is restricted to only those entities and relations explicitly mentioned in $e'$. Within this smaller, discrete evidence-driven world, the evidence $e'$ becomes formally complete, which provides two key benefits: it makes the computation tractable and elegantly handles missing data in a principled manner.

Crucially, this method yields simple, analytical expressions for inductive inference by applying the characteristic function derived in the previous section. For example, the predictive probability that the next observed individual, $a_{n+1}$, is of a specific kind (attributive constituent) $Ct_i$, given that this kind has been observed $n_i$ times, simplifies to:
$$p(Ct_i(a_{n+1})|e') = \frac{n_i + \lambda/K'}{n + \lambda}$$
where $K'$ is the number of possible kinds in the sub-language $\mathcal{L}'$. This approach is supported by rigorous theoretical guarantees establishing its convergence and sample complexity, marking a key advantage of Hintikka's system over earlier formulations, particularly in its robust handling of universal laws \cite{NIINILUOTO2011311}.

\section{A Quantitative Framework for Semantic Information}

\subsection{The Classical Foundation and Its Limits}
The seminal work by Carnap and Bar-Hillel provides the classical foundation for semantic information. Their central idea is that the information in a formula $m$ is proportional to the set of possible states of the world it logically excludes. This is quantified by the \textit{content measure}, which is based on the inductive probability $q_e(m)$:

\begin{definition}[Carnap's Content Measure]
The semantic content of a FOL formula $m$ given evidence $e$ is defined as:
\begin{equation}
    \text{cont}_e(m) \triangleq 1 - q_e(m)
\end{equation}
\end{definition}

A message is more informative if it is less probable, as it eliminates a larger fraction of possible states. However, this intuitive measure is not additive. For two \textit{inductively independent} FOL formulas $m_1$ and $m_2$ (where $q_e(m_1 \land m_2) = q_e(m_1)q_e(m_2)$), the total information is not the sum of the parts: $\text{cont}_e(m_1 \land m_2) \neq \text{cont}_e(m_1) + \text{cont}_e(m_2)$. This led Carnap and Bar-Hillel to introduce a second measure, $\text{inf}_e(m) \triangleq -\log_2 q_e(m)$, which is additive. This duality between a content-based and a rarity-based measure has been a long-standing issue in semantic information theory.

\vspace{-5pt}
\subsection{A Bifurcated Theory of Semantic Information}
We propose to resolve this duality by positing that it reflects two distinct, complementary levels of information. Our analysis assumes static information, ignoring time-variant (a.k.a., temporal) components. We distinguish between the information related to the \textbf{signifier} (the symbol being transmitted) and the \textbf{signified} (the concept or meaning it points to).

\subsubsection{Level 1: Physical Entropy of the Signifier}
This level addresses the statistical complexity of encoding the signifiers (our semantic symbols) for transmission. It quantifies the average number of bits required to represent a symbol from a given semantic alphabet $\mathcal{A}$. This is a direct application of Shannon's entropy formula using our inductive probabilities.
\begin{definition}[Physical Semantic Entropy]
The physical entropy of a semantic source $\mathbf{f}$ with a FOL-based semantic alphabet $\mathcal{A}$ given evidence $e$ is:
\begin{equation}
    H_{\text{phys}}(\mathbf{f}|e) \triangleq -\sum_{A_i \in \mathcal{A}} q_e(A_i) \log_2 q_e(A_i)
\end{equation}
\end{definition}
The alphabet $\mathcal{A}$ can be defined at different levels of abstraction (e.g., individual predicates and entities, or entire constituents). As the evidence $|e| \to \infty$, the inductive probabilities $q_e(A_i)$ converge to the true frequencies, and thus $H_{\text{phys}}$ converges to the classical Shannon entropy.

\subsubsection{Level 2: Calibrated Content of the Signified}
This level addresses the meaning itself: "How much does a message reduce uncertainty about the true state of the world?" Classical measures fail for infinite universes because their normalization depends on the number of entities. Our framework, built on Hintikka's system, resolves this by grounding information in the finite set of possible \textbf{kinds} of individuals, not the potentially infinite set of individuals themselves.

\begin{definition}[Language Granularity]
For a dyadic language with $m$ basic predicates, there are $K_Q = 2^{2m}$ unique Q-predicates. The \textbf{granularity} of the language, $G$, is the total number of distinct individual kinds (attributive constituents) it can specify:
\begin{equation}
    G \triangleq 2^{K_Q} = 2^{2^{2m}}
\end{equation}
\end{definition}

This granularity $G$ is finite and independent of the number of entities in the universe, representing the true "volume" of the logical space.

\subsection{Axiomatic Foundation of the Content Measure}
We derive our universal measure of semantic content from a set of fundamental axioms that formalize how information quantifies the reduction of uncertainty.

\begin{axiom}[Linearity in Elimination]
The semantic information conveyed by a message $\mathbf{m}$ must be directly proportional to the fraction of the logical state space it eliminates, as measured by $\text{cont}_e(\mathbf{m})$.
\begin{equation}
    H(\mathbf{m}) \propto \text{cont}_e(\mathbf{m})
\end{equation}
\end{axiom}
\vspace{-8pt}
\begin{axiom}[Calibration by Granularity]
To ensure comparability across systems with different descriptive capacities, the semantic information must be directly proportional to the total logical volume of the universe, quantified by the language granularity, $G$.
\begin{equation}
    H(\mathbf{m}) \propto G
\end{equation}
\end{axiom}
\vspace{-5pt}
Combining these principles provides a complete characterization of the calibrated information content for a single message $\mathbf{m}$.

\begin{definition}[Calibrated Content-Information]
The calibrated content-information of a single message $\mathbf{m}$ is jointly proportional to the language granularity $G$ and the fraction of the logical space it eliminates (with the constant of proportionality set to 1):
\begin{equation}
    H(\mathbf{m}) \triangleq G \cdot \text{cont}_e(\mathbf{m})
\end{equation}
\end{definition}
\vspace{-5pt}
Since communication is a probabilistic process, we are interested in the \textit{average} information conveyed by a source. This is found by taking the expectation over all possible messages in a logical partition $\mathcal{M}$ (e.g., the set of all constituents $\mathcal{C}$).

\begin{proposition}[Average Semantic Content-Information]
Given a logical partition $\mathcal{M}=\{m_i\}$ of a universe with granularity $G$, the average content-information of a source given evidence $e$ is:
\begin{align}
    H_{\text{cont}}(\mathcal{M}|e) & \triangleq \mathbb{E}_{m_i \sim q_e(m_i)}[H(m_i)] \nonumber \\
    & = \sum_{m_i \in \mathcal{M}} q_e(m_i) \left[ G \cdot \text{cont}_e(m_i) \right]
\end{align}
\end{proposition}

This axiomatic derivation provides a firm theoretical grounding for our proposed semantic compression objectives.
\vspace{-15pt}
\subsection{Semantic Compression Objectives}
These measures form the basis for a semantic rate-distortion theory, where the goal is to create a compressed representation $\hat{s}$ of an original source $s$ that minimizes rate while maximizing retained semantic meaning.
\begin{theorem}[Semantic Source Coding Objectives]
Let $f$ be a semantic source message, and $\hat{c}$ its semantically compressed representation. Let the communication rate be constrained by a budget of $R^* > 0$ bits. Let $q_e(\hat{c}|f)$ be a probabilistic semantic encoding channel, and $q_e(C_w|h)$ be the inductive posterior over world constituents given a FOL statement $h$. The lossy semantic compression schemes are then solution to one of the following optimization problems:
\begin{enumerate}
    \item \textbf{Maximization of Average Content Fidelity:} The optimal semantic encoding that maximizes the average semantic content-information between the semantic source \textbf{f}, its semantically compressed representation $c^*$ is given by:
    \begin{equation}
        c^* = \argmax_{c \; \sim \; q_e(f)} H_{\text{cont}}(\mathbf{f}) \quad \text{s.t.} \quad H_{\text{phys}}(c) \leq R^* \text{ bits}
    \end{equation}
    \vspace{-8pt}
    \item \textbf{Maximization of Instantaneous Content:} For a given candidate set of FOL formulas $\mathcal{I}_s$ at the transmitter, the optimal formula $s^* \in \mathcal{I}_s$ to transmit is the one whose deterministic encoding $g(s^*)=\hat{s}^*$ yields the highest semantic content, subject to a hard bit-length constraint:
    \begin{equation}
        s^* = \argmax_{s \in \mathcal{I}_s} \;\; \text{cont}_e(g(s)) \quad \text{s.t.} \quad H_{\text{phys}}(c) \leq R^* \text{ bits}
    \end{equation}
\end{enumerate}
\end{theorem}

\vspace{-20pt}
\section{Experimental Validation}
\vspace{-2pt}
To demonstrate the practical utility of our proposed framework, we conducted a series of experiments on a semantic text transmission task and LogiCity \cite{Li2024LogiCityAN} driving simulator.
\vspace{-2pt}
\subsection{Experimental Setup and Results}
Experiments used a custom FOL dataset\footnote{https://github.com/xiongsiheng/Inductive-Semantic-Communication-Dataset} of seven stories, each as an independent semantic source. Texts were converted to dyadic FOL representations for computing semantic measures, with Lempel–Ziv (Gzip) compression as the baseline.

\textbf{Source Content Analysis:}  
We first evaluated the calibrated content-entropy $H_{\text{cont}}$ for each story (Table~\ref{table:cont_entropic}), quantifying total logical information and its normalized efficiency $H_{\text{cont}} / G$. Results show large variation in informativeness across sources, with Story 1 exhibiting the highest logical content and Story 2 the lowest, highlighting the scalability of our framework for measuring semantic volume over large logical spaces.

\textbf{Physical–Semantic Compression:}  
Next, we assessed compression based on physical semantic entropy $H_{\text{phys}}$ (Eq.~4, Table~\ref{table:lossless}). The two-stage pipeline—structural reduction via FOL conversion followed by entropy coding using inductive probabilities $q_e(A_i)$—achieved compression ratios of $8.35\times$–$15.54\times$, significantly surpassing direct Gzip performance.

\begin{figure}[h]
\centering
\includegraphics[width=0.7\linewidth]{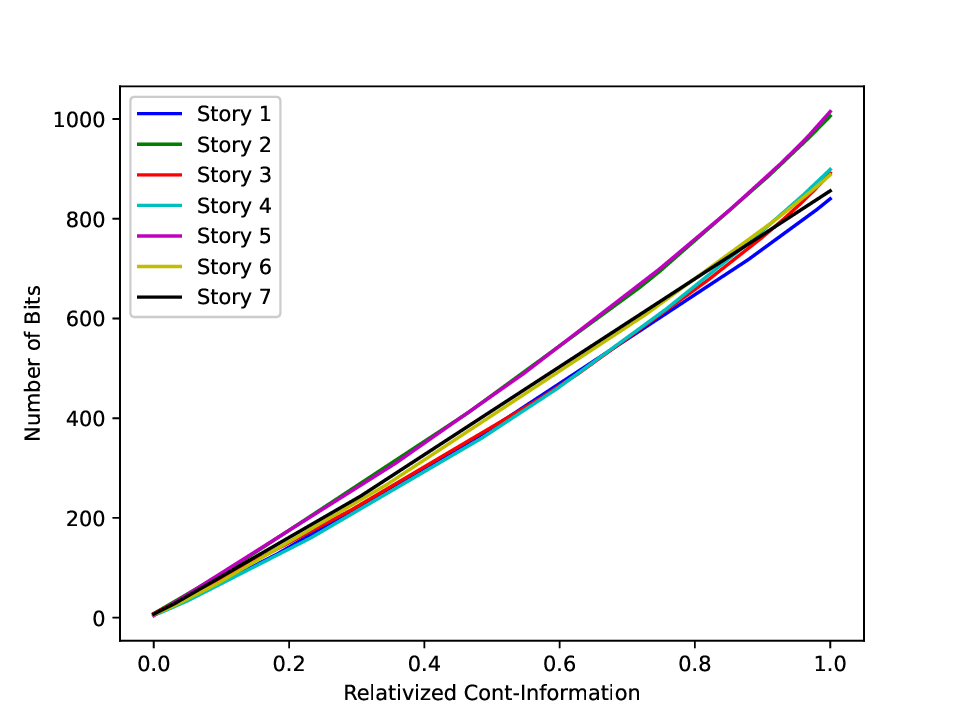}
\caption{Content-Semantic Source Coding}
\vspace{-15pt}
\label{fig:my_label}
\end{figure}

\textbf{Content–Semantic Compression:}  
Lossy compression was evaluated under our semantic rate–distortion objective (Theorem~1.1), optimizing average content fidelity under rate constraints. Figure~\ref{fig:my_label} illustrates the trade-off between normalized mutual content-information $I_{\text{cont}}(s;\hat{s})/H_{\text{cont}}(s)$ and transmission rate. Results demonstrate an explicit and tunable balance between semantic fidelity and bit cost, confirming the framework’s ability to guide compression toward meaning-preserving transmission.

Second set of experiments are conducted over LogiCity simulator to demonstrate improvements in deduction performance.

\textbf{Deduction in the LogiCity Environment:}  
\textit{LogiCity} is a $250 \times 250$ grid-based urban simulation comprising 25 building blocks connected by bidirectional traffic and walking streets. The expert configuration instantiates $N=18$ heterogeneous agents, including vehicles and pedestrians with diverse semantic attributes (e.g., ambulances, buses, tiro and reckless drivers, police, and pedestrians). Agents are assigned collision-free start and goal positions and follow globally planned A* paths. Each agent observes a limited field of view (FOV) with radius $r$, producing local observations and a set of visible agents inside the FOV. Predicate groundings evaluate semantic and spatial relations (e.g., \texttt{IsAmbulance}, \texttt{IsOld}, \texttt{IsClose}, \texttt{LeftOf}), which are processed by a Z3 SMT solver under first-order logic rules to deduce if the current state is compliant with traffic constraints.

To enhance deduction performance, a \textit{Global Navigation Assistant} (GNA) aggregates global state information, comprising of entirety of observations transmitted to GNA by local agents, and broadcasts a subset of $K$ FOL expressions to all agents. The rule base comprises 13 disjunctive FOL-based traffic rules. Two broadcast strategies are evaluated for $K \in \{1,\ldots,10\}$: (1) \textbf{cont-based} selection, as per Eq.~(11), and (2) \textbf{random} selection, which uniformly samples $K$ expressions. Each configuration is tested over five trials of 100 timesteps. The evaluation metrics are: (i) average percent of fully observed / partially observed / unobserved rules, and (ii) percent informativeness, measuring how well the agent’s information explicates the true world state. Results reveal a fundamental trade-off: cont-based selection maximizes information density by focusing on high-criticality predicates, while random selection increases entity-type diversity, improving rule coverage and deductive completeness.

Table~\ref{table:gna_comparison_normalized_percent} confirms that cont-based selection consistently outperforms random selection across most $K$ values. For example, at $K=3$, cont-based achieves 50.9\% fully observed rules and 50.4\% explicative informativeness, exceeding random selection (36.9\% and 41.5\%, respectively). As $K$ increases, cont-based maintains superior performance, reaching 81.4\% explicative informativeness at $K=10$, compared to 62.5\% for random. The marginal advantage of random selection at midrange $K$ (e.g., $K=7$–$8$) at rule observations arises from diversity of observations, i.e., occasional exposure to rare entity types that satisfy otherwise undersampled. However, these gains are inconsistent and diminish as semantic coverage expands. Overall, cont-based broadcasting delivers higher logical completeness and information efficiency, validating that prioritizing high-content predicates maximizes deductive fidelity under bandwidth constraints.

\vspace{-6pt}
\section{Conclusion}
\vspace{-5pt}
This paper introduced a computable framework for semantic communication that separates two information levels: \textbf{physical entropy} ($H_{\text{phys}}$) for symbol transmission and \textbf{content-entropy} ($H_{\text{cont}}$), a calibrated measure of informativeness grounded in inductive logic. Calibrated by language granularity, it allows consistent comparison of content-informativeness across logical systems. Building on this, we developed a semantic rate–distortion theory linking transmission rate to semantic fidelity. Experiments verified its utility, showing a clear trade-off between bit cost and preserved content, with improved agent-level world understanding and deduction accuracy. The framework provides an interpretable, mathematically grounded alternative to black-box communication for inference, advancing semantic communication toward a rigorous quantitative science.

\vspace{-2pt}
\begin{table}[h] 
\centering
\begin{tabular}{|c|c|c|c|}
\hline
\textbf{ID} & \textbf{$H_{\text{cont}_e\text{, calib}}$}& \textbf{$H_{\text{cont}_e\text{, min}}$} & \textbf{$H_{\text{cont}_e\text{, max}}$}\\
\hline
1 &1.18$\times$$10^{-563}$ &9.0$\times$$10^{14161}$ &1.0\\
\hline
2 &1.31$\times$$10^{-14725}$ &1.0 &1.1$\times$$10^{-14162}$\\
\hline
3 &4.34$\times$$10^{-9459}$ &3.3$\times$$10^{5266}$ &3.7$\times$$10^{-8896}$\\
\hline
4 &1.08$\times$$10^{-867}$ &8.2$\times$$10^{13857}$ &9.1$\times$$10^{-305}$\\
\hline
5 &2.54$\times$$10^{-12274}$ &1.9$\times$$10^{2451}$ &2.1$\times$$10^{-11711}$\\
\hline
6 &1.20$\times$$10^{-619}$ &9.2$\times$$10^{14105}$ &1.0$\times$$10^{-56}$\\
\hline
7 &1.02$\times$$10^{-2802}$ &7.9$\times$$10^{11922}$ &8.7$\times$$10^{-2240}$\\
\hline
\end{tabular}
\vspace{3mm} 
\caption{Comparison of Calibrated, Min-Normalized, Max-Normalized Cont-Entropic Measures of Rival Stories}
\label{table:cont_entropic}
\end{table}

\vspace{-6.8mm} 

\begin{table}[h] 
\centering
\begin{tabular}{|c|c|c|}
\hline
\textbf{Story ID} & \textbf{Lossless Semantic Compr.} & \textbf{Shannon Compr.}\\
\hline
Story 1 & 840 bits & 11018 bits\\
\hline
Story 2 & 1006 bits & 13116 bits\\
\hline
Story 3 & 892 bits & 13862 bits\\
\hline
Story 4 & 899 bits & 11713 bits\\
\hline
Story 5 & 1015 bits & 8462 bits\\
\hline
Story 6 & 888 bits & 12269 bits\\
\hline
Story 7 & 856 bits & 11686 bits\\
\hline
\end{tabular}
\vspace{3mm} 
\caption{Comparison of Physical-Semantic and Shannon Compression}
\label{table:lossless}
\end{table}

\vspace{-11mm} 

\begin{table}[h]
\vspace{3mm}
\centering
\begin{tabular}{|c|c|c|c|c|}
\hline
\textbf{$k$} & \textbf{Full (\%)} & \textbf{Partial (\%)} & \textbf{Unobs. (\%)} & \textbf{Explication (\%)}\\
\hline
0  & 18.5 & 43.2 & 38.4 & 33.1\\
\hline
1  & 33.7 / 23.9 & 40.2 / 44.5 & 26.3 / 31.7 & 38.2 / 35.7\\
\hline
2  & 33.7 / 30.5 & 40.2 / 43.8 & 26.3 / 25.8 & 44.4 / 38.1\\
\hline
3  & 50.9 / 36.9 & 22.9 / 42.9 & 26.2 / 20.2 & 50.4 / 41.5\\
\hline
4  & 50.9 / 44.8 & 23.1 / 39.5 & 26.1 / 15.8 & 56.5 / 43.9\\
\hline
5  & 51.1 / 50.5 & 23.0 / 37.0 & 25.9 / 12.6 & 62.0 / 46.6\\
\hline
6  & 57.9 / 55.8 & 22.9 / 34.5 & 19.2 / 9.8  & 66.8 / 49.6\\
\hline
7  & 61.7 / 62.7 & 19.2 / 29.9 & 19.1 / 7.3  & 71.0 / 53.1\\
\hline
8  & 62.2 / 68.7 & 19.2 / 25.5 & 18.7 / 5.8  & 74.7 / 56.6\\
\hline
9  & 61.5 / 72.5 & 19.2 / 23.3 & 19.2 / 4.2  & 78.2 / 59.7\\
\hline
10 & 65.7 / 78.4 & 20.2 / 18.7 & 14.1 / 2.9  & 81.4 / 62.5\\
\hline
\end{tabular}
\vspace{3mm}
\caption{Comparison of Top-$k$ broadcast from RSU based on Semantic and Random selection (Format: Semantic / Random).}
\label{table:gna_comparison_normalized_percent}
\end{table}

\vspace{-10pt}

\vspace{-10pt} 

\bibliographystyle{unsrt} 
\bibliography{refs} 

\end{document}